\documentstyle[preprint,aps]{revtex}
\begin{document}

\draft

\title{$J/{\Psi}$ suppressions in a thermally equilibrating \\
quark-gluon plasma at RHIC}

\author{Gouranga C Nayak \thanks{ e-mail:gcn@iitk.ernet.in}
}
\address{Department of Physics, Indian Institute of 
technology, Kanpur -- 208 016, INDIA }

\maketitle

\begin{abstract}
We estimate the survival probability of $J/{\psi}$
in relativistic heavy-ion
collisions, using both short-distance QCD
and nuclear absorption mechanism.
The suppression is found to be almost 100 percent
for a thermally equilibrating quark-gluon plasma.
The measurement of such a huge
suppression of $J/{\psi}$ may suggest
the existence of a deconfined partonic medium, 
possibly a thermalised quark-gluon plasma, at RHIC.

\end{abstract}

\vspace{1.5 cm}


\section{INTRODUCTION}

Recently a lot of effort is being made to 
detect a new phase of matter, namely 
quark-gluon-plasma(QGP).
It is known from lattice quantum chromodynamics 
\cite{karsch} that hadronic matter
at sufficiently high temperatures
($\sim 200 MeV$) and densities
undergoes a transition to this quark-gluon plasma phase.
While such a phase surely did exist in the early 
universe, it is interesting if we can recreate the early
universe and experimentally verify this QCD phase in the
laboratory {\it i.e.}
in ultra relativistic heavy ion collisions (URHIC).
In the near future, the 
relativistic heavy ion colliders(RHIC) at BNL and 
large Hadron Colliders(LHC) at CERN \cite{awes,gust}
will be able to study such a phase. 
However this deconfined phase is not accesible directly at 
these collider
experiments. The signatures are therefore necessarily indirect,
and the prominent ones are 
: 1) $J/{\psi}$ suppression \cite{matsui},
2) electromagnetic probes such as dilepton and direct photon production
\cite{strickland,shuryak,sinha}, and 3) strangeness enhancement
\cite{rafelski}.

Dileptons and single photons have long been proposed as useful
probes of the plasma \cite{sinha}, as once produced,
they hardly interact with the strongly interacting matter 
and thus carry the details of the circumstances
of their production. However, as 
they are also produced via hadronic decays
in the later stages of evolution, where plasma expands and cools,
it becomes difficult to distinguish between them.
In that sense the production of $c\bar{c}$ is a clean process, 
as it is not produced in the later stages due to its heavy mass.
The production of c$\bar{c}$ takes place mainly
at the hard vertex. Once produced, $c\bar{c}$ will evolve 
in to open D mesons and charmonium states, such as $J/{\psi}$,
$\chi$, etc., while travelling through
this dense phase of matter. Unlike hadronic collisions,
where an enhancement of $J/{\psi}$ is found
at tevatron energy \cite{ge3},
both $c\bar{c}$ and $J/{\psi}$ get supressed while traveling
through different stages of quark-gluon plasma. There might be 
thermal charm productions in the very early stage of
the plasma, where the temperature is very high, but no production
of c$\bar{c}$ can occur in the later stages due to decrease in
the temperature. 

The various stages by which the complete evolution of 
quark-gluon plasma is descrbed in URHIC are, i) 
pre-equilibrium, ii) equilibrium, where one actually studies
the thermalised quark-gluon plasma, iii)
cooling and iv) hadronisation.
The production of an equilibrated quark-gluon plasma 
in stage (ii) crucially
depends on the pre-equilibrium evolution, {\it i.e.} on stage
(i). Matsui and Satz \cite{matsui} has suggested the suppression
of $J/{\psi}$
in the equilibrium phase. In their study,
the debye screening length which is calculated from lattice QCD, 
is found to be less than the $J/{\psi}$ radius.
This forbids the binding of c$\bar{c}$ to $J/{\psi}$ 
in the equilibrated quark-gluon
plasma. However, such a study is not available for 
an equilibrating quark-gluon plasma.
There could be any interesting effects
in the pre-equilibrium stage, when c$\bar{c}$ or
$J/{\psi}$ travel before reaching
the equilibrium stage. As high energy 
deconfined partons are 
also present in the pre-equilibrium stage, $J/{\psi}$ is supressed
due to its interaction with these deconfined partons. Since the
$J/{\psi}$ dissociation cross section inside a
deconfined partons is very different from that inside a 
hadronic gas (see section II), 
the study of $J/{\psi}$ suppression can then provide us
with evidence for color deconfinement in the parton plasma and
possibly with information on the QCD phase transition.

Another outcome of the pre-equilibrium study is the 
prediction of the equilibration time, the time
at which quark-gluon plasma equilibrates. 
A detail knowledge of equilibration time and $J/{\psi}$ formation time
is very crucial in determining the $J/{\psi}$ suppressions in URHIC.
This is because, a large equilibration time may permit
a $J/{\psi}$ formation from c$\bar{c}$ pairs
before plasma equilibrates. 
Then the interaction of a fully formed $J/{\psi}$ with deconfined 
partons is prefered than screening.
If the $J/{\psi}$ formation time is greater than the equilibration
time or comparable to it, then c$\bar{c}$ interacts 
with the deconfined partons before screening starts operating
in the equilibrium stage.
The interaction of a c$\bar{c}$ with the partons before plasma 
equilibriates is not studied yet, and such a calculation 
is beyond the scope of this paper.

In any case, a careful study of pre-equilibrium
evolution of quark-gluon plasma is the consistent way to study
production and equilibration of quark-gluon plasma in URHIC.
From this point of view it is necessary to study what happens actually
to the $J/{\psi}$ suppressions at different stages of evolution of QGP,
rather than estimating it in an equilibrated quark-gluon plasma.
We study here the suppression of a fully formed $J/{\psi}$
in equilibrating quark-gluon plasma using short distance QCD
and suppression of a nascent $J/{\psi}$ by nuclear absorption.
The $J/{\psi}$ may also be supressed in later
stages due to inelastic collisions with hadrons, but we do not
consider it here. This suppressions will be less due to lesser
value of the hadron-$J/{\psi}$ inelastic cross section (described
in section II).

However, $J/{\psi}$ suppression is also measured in p-A collisions
where there is no existence of quark-gluon plasma \cite{sps,na38}.
At these experiments the suppression of $J/{\psi}$ is 
due to the presence of a nuclear medium.
The prominent mechanism by which these datas are explained at SPS
is via nuclear absorption, the
suppression of a nascent $J/{\psi}$ 
before it forms a physical resonance
\cite{huffn,kharezeev}.
Within this mechanism, one does not assume the existence of
a quark-gluon plasma phase or a highly densed
deconfined partonic medium.
This is the case with some light nuclei collisions at SPS \cite{sps,na38}.
However the recent measurements of $J/{\psi}$ suppression
by NA50 collaboration \cite{na50} 
(Pb-Pb collisions at $\sqrt{s}= 17 GeV$)
yeild an excess suppression of $J/{\psi}$, which
is not explained by the above conventional approach.
Presence of a nuclear medium may not be enough to explain this
data. There are 
speculations that this suppression is due to the existence of a
deconfined partonic medium \cite{satz2} or a high energy density matter
\cite{bla}. 

Whether an equlibrated plasma has formed in Pb-Pb collisons at
SPS($\sqrt{s} = 17 GeV$) or not, is debatable, but at 
RHIC(Au-Au collisions at $\sqrt{s}= 200 GeV$) we may go closer to the
equilibrated quark-gluon plasma. In this paper we estimate the 
$J/{\psi}$
suppressions at RHIC and its possible implication on deconfinement. 

However before studying charmonium suppressions in 
quark-gluon plasma it is essential to study
the corresponding situations in hadronic collisions.
The experimental analysis of 
quarkonia production in high energy
hadronic collisions at tevatron energy
\cite{ge3} revealed a drastic disagreement 
with the predictions from color singlet model \cite{bair}.
A substantial fraction of quarkonium production
at very high energy arises from intermediate
color octet fluctuation during the evolution
into final state color singlet Q$\bar{Q}$ pair.
Now a rigorous theoretical foundation has been 
achived by Bodwin, Braaten,
and Leepage \cite{bod}, who devloped a 
QCD formalism by marrying pQCD 
and an effective field theory within non-relativistic QCD(NRQCD).
This formalism accounts
for the production of both color singlet
and color octet $c\bar{c}$ states, that
evolve in to final color singlet quarkonium. Within NRQCD, 
the production
amplitude is expanded in powers of both the strong
coupling $\alpha_s$ and the velocity of heavy quark($v^2 = .23$ for
c$\bar{c}$ system), which 
includes higher fock-state components. The long 
distance non-pertubative matrix element is either fitted from
experiments or taken, in principle, 
from lattice QCD\cite{150}. This technique has been used
to study many heavy quarkonium production processes in hadronic 
collisons. 
However a space time evolution of quarkonium using this
technique has not been achived.
This is crucial for studying the evolution of $J/{\psi}$
in nuclear collisions. The existence of background
chromoelectric field \cite{nayak} and high dense partonic medium
complicates the study of the evolution of c$\bar{c}$ to $J/{\psi}$,
both at RHIC and LHC.
As c$\bar{c}$ takes
$\simeq$ 1.0 fm (time in c$\bar{c}$ rest frame)
to form a $J/{\psi}$ bound state, it 
interacts with the nearby light partons and with
the background chromoelectric field during this period.
No attempts is made on this line to study $J/{\psi}$ production
and its subsequent evolution in nucleus-nucleus collisions. 
This might be the consistent way to study $J/{\psi}$ suppression
in URHIC.

Since we study the suppression of $J/{\psi}$ in equilibrating
quark-gluon plasma,
it is relevant to discuss the process of equilibration
of QGP in URHIC.
The equilibration of QGP in heavy-ion collisions
is not completely studied yet. 
The rate of equilibration is different for diiferent models.
Some relevant models that describe the equilibration of QGP 
are parton cascade model(PCM) \cite{geiger},
heavy-ion jet interaction generator(HIJING) \cite{wang} and
color flux-tube model \cite{nayak}. PCM and HIJING are both
perturbative QCD based models, and color flux-tube model
is a field theoritical model. The former models describe the
evolution of hard and semihard partons and the later
describes the production of soft partons via non-perturbative
Schwinger mechanism. For a complete study of production
and equilibration of quark-gluon plasma, color flux-tube
model may be combined with the pQCD based
models. As an initial attempt,
we use here the distribution of partons from parton cascade
model to study $J/{\psi}$ suppressions in a thermally
equilibrating quark-gluon plasma at RHIC. 
Within parton cascade model, a relativistic 
transport equation is solved to study the distribution of
partons by considering the direct collision 
processes $2->2$, the inelastic collision processes
$2->1$, and the decay processes $1->2$ \cite{geiger}.
More explicitely, this is written as
\begin{equation}
p^{\mu} \partial_{\mu} f_i(x,p) = C_i(x,p)
\end{equation}
where $f_i(x,p)$ is the distribution of a particular type of parton
$i$ in the usual phase space. The collisions term $C_i(x,p)$ is derived
from pQCD taking all the processes mentioned above.
The initial condition on $f_i(x,p)$
is the distribution of partons inside the nucleus
before the two nuclei collide with each other \cite{geiger3}.
However for a complete study, the addition of soft partons 
and the effect of background field has to be taken into account. These
backgruond electric field may not exit in the very early stage where
pQCD is applicable \cite{eskola}, but in later times this will
have an important role in the equilibration of the plasma
\cite{eskola,eskola1}.
Such calculations 
will be available once the color flux-tube model is combined
to the pQCD based models \cite{nayak1}. The addition
of soft partons to hard and semi hard partons
within color flux-tube model will enhance 
the $J/{\psi}$ suppression (see below).
This will be taken up separately.

In section {\bf II} we describe the short-distance QCD and
the dissociation of $J/{\psi}$. Section {\bf III}
contains a brief discussion of nuclear absorpton mechanism.
Results and discussions are presented in section {\bf IV}, for RHIC
energy. We summarise and conclude the main results in section {\bf V}.

\section{$J/{\psi}$ dissociation by deconfined gluons}

In the framework developed by Bhanot and Peskin \cite{peskin,bhanot},
interactions between light
hadrons and deeply bound, heavy quarkonium states, such as the
$J/\psi$, are mediated by short-range color dipole interactions. 
For sufficiently heavy quarks, the dissociation of quarkonium 
states by interaction with light hadrons is fully
accounted for by short-distance QCD \cite{peskin,bhanot}.
Because of its small size, a heavy quarkonium
can probe the short distance properties of light hadron.
A parton based calculation of the $J/{\psi}$-hadron cross section
is thus possible via an operator product expansion method, 
simillar to that used in deeply inelastic lepton-hadron
scatterings \cite{peskin,bhanot,kaidalov,kha}.
These pertubative calculations
become valid when the space and time scale associated
with the quarkonium state, $r_Q$ and $t_Q$, are small in comparison
to the nonperturbative scale $\lambda_{QCD}^{-1}$, which is the 
characteristic size of the light hadrons, {\it i.e.}
$r_Q <<< \lambda_{QCD}^{-1}$, and $t_Q <<< \lambda_{QCD}^{-1}$.
For $J/{\psi}$ ground state $r_{J/{\psi}} \simeq 0.2 fm = (1 GeV)^{-1}$
and $E_{J/{\psi}}$(${2 M_D - M_{{\psi}^\prime}}$) $\simeq$ 0.64 GeV.
With $\lambda_{QCD} = 0.2 GeV$, the above inequalities seem to be well
satisfied \cite{hwa2} and one expect that the dissociation 
of $J/{\psi}$ by hadron will be governed by the 
$J/{\psi}$-hadron break-up cross section as calculated
in short distance QCD. The operator 
product expansion allows one to express the hadron-$J/{\psi}$ 
inelastic cross section in terms of the convolution of the 
gluon-$J/{\psi}$ dissociation cross section with the gluon distribution
inside the hadron. 
The gluon-$J/{\psi}$ dissociation cross
section is given by \cite{peskin,bhanot,kha}:
\begin{equation}
\sigma(q^0) = {{2 \pi } \over {3}} {(32/3)^2}{({{16 \pi}
\over {3 g_s^2}})}{(1/{m_Q^2})}{{(q^0/{\epsilon_0} -
1)^{3/2}} \over {({q^0/{\epsilon_0}})^5}}.
\end{equation}
Here $g_s$ is the coupling between gluon and charm quark, $m_Q$ the 
charm quark mass, and $q^0$ the gluon energy in
the $J/{\psi}$ rest frame.
The heavy quarkonium is coulomb like, and one writes,
{$\epsilon_0 = {({{3 g_s^2 } \over {16 \pi}})^2 m_Q}$} 
\cite{peskin,kha}.

The gluons which are soft inside a pion are not capable
of dissociating a charmonium. Hence they give a low
$\pi$-${J/{\psi}}$ cross-section.
However, deconfined gluons, which carry enough energies, are sufficient
to break a charmonium \cite{hwa2}. 
These high energy partons are present in URHIC, as observed in
several calculations, such as \cite{nayak,geiger}.
These deconfined gluons can break a fully formed $J/{\psi}$ that exit
inside QGP. This conclusion does not seem to be
affected substantially by nonperturbative effects \cite{mclerran}.

We use the above g-$J/{\psi}$ dissociation cross section to study the
$J/{\psi}$ suppressions in a thermally equilibrating quark-gluon 
plasma at RHIC.
For the central collisions, the $J/{\psi}$ survival probability is:
\begin{equation}
S_{g-{J/{\psi}}}(P_T) =
\exp[ - \int d \tau n_g(\tau) <v_{rel} \sigma(k \cdot v)>_k].
\end{equation}
Here $n_g(\tau)$ is the gluon number density which
evolves as system expands longitudinally. According to Bjorken
scenerio \cite{bjorken},
this number density is a function of the boost invariant
parameter $\tau$ ($\tau = \sqrt{(t^2 - z^2)}$).
The thermal average g-$J/{\psi}$ 
cross section, $<v_{rel} \sigma(k \cdot v)>_k$, is
\begin{equation}
<v_{rel} \sigma(k \cdot v)>_k =
{{\int d^3k v_{rel} \sigma(k \cdot v) f(k^0, T(\tau))} \over 
{\int d^3k f(k^0, T(\tau))}}.
\end{equation}
Here $v (\equiv (M_T, \vec{P_T}, 0)/{M_{J/{\psi}}})$
is the four-velocity
of $J/{\psi}$ in central rapidity region and k is the 
four-momentum of gluon in the parton gas.
In the thermal average cross-section 
we use the distribution function,  
$f(k^0, T(\tau)) = {{a(\tau)} \over {{\exp({{k^0}/{T(\tau)}}) -1}}}$
for gluon.
Here $a(\tau)$ captures the deviation from equilibrium,
which in principle, is determined from the relation,
$n_g{(\tau)}=\int d\Gamma (p^\mu u_\mu) f(p,T(\tau))$,
by knowing the evolution of temperature  and $n_g$, for an 
equilibrating quark-gluon plasma.
Here ${d{\Gamma} = {{\gamma d^3 {p}}
\over {{({2 \pi})}^3 {p_0}}}}$ with $\gamma= 2 \times 
8$(the product of spin and color degeneracy),
and u( = (t/{$\tau$}, 0, 0, z/{$\tau$})) is the flow velocity.
However, as $a(\tau)$ cancels
both from numerator and denominator of equation(4),
we need not know its form.
What matters is $n_g(\tau)$ and T($\tau$) whose 
evolutions are taken from PCM \cite{geiger}.
Within PCM, $n_g(\tau) \simeq$ 0.72$n(\tau)$, with
$n(\tau) = 565 fm^{-3}({{\tau} \over {\tau_0}})^{-0.90}$
and $T(\tau) = 950 MeV ({{\tau} \over {\tau_0}})^{-0.30}$, for
a thermally equilibrating quark-gluon plasma at RHIC.
Here $\tau_0 = .05 fm$. Using this distributions
the integration in equation (3)
is performed numerically to study the $J/{\psi}$ survival
probability.

\section{$J/{\psi}$ suppression by nuclear absorption}

Consider p-A collisions:
unlike hadronic collisions,
where an enhancement of $J/{\psi}$ is
measured at very high energy \cite{ge3},
$J/{\psi}$ suppression is measured in p-A collisions. A number of
explanations have been given \cite{huffn,kharezeev,won,brod}
for this suppression, among which nuclear absorpton is the prominent.
This nuclear absorption 
is well described according to Glauber theory 
\cite{huffn}. Within this theory a 
$J/{\psi}$ survival probability is written as
\begin{equation}
S_{pA} = {1 \over {A}} \int d^2 b dz \rho_A({\bf{b}},z)
\exp(- \int_z^{\infty} d z^{\prime} \rho_A({\bf{b}},z^{\prime})
\sigma_{abs} (1 - {1/A})).
\end{equation}
Here ${\bf{b}}$ is the impact parameter 
where the proton collides with
the nucleus and z is the longitudinal distance at which 
a $c\bar{c}$ pair is created. After the
$c\bar{c}$ is produced, it travels in the forward direction 
colliding inelastically with the surrounding nucleons.
The inelastic c$\bar{c}$-N absorption cross section, 
$\sigma_{abs}$, 
is determined from the experiment. The nuclear 
density,
$\rho_A$ at a point ${\bf{r}}({\bf{b}},z)$ is normalised such 
that: $\int d^3 r \rho_A({\bf{r}})$ = A. 
After integrating over z, we get  
\begin{equation}
S_{pA} = ({1 \over {A^{\prime} \sigma_{abs}}}) \int d^2 b
{[1 - {\exp(- \sigma_{abs} (1 - {1/A}) T_A(\bf{b}))}]}, 
\end{equation}
where $ T_A({\bf{s}}) =
\int_{- \infty}^{+ \infty} \rho_A({\bf{s}}, z) dz$, is
the nucleon density per unit area
in the transverse plane(transverse to the collision
axis), and $A^{\prime} = A(1 - {1/A})$.
Extendng to nucleus-nucleus collisions \cite{huffn} 
we write the survival probability as,
\begin{eqnarray}
S_{AB} = {1 \over {A^{\prime} B^{\prime}
\sigma_{a}^2}} \int d^2 b \int d^2 s \left[ 1 - {\exp(-
\sigma_{abs} (1 - {1/A}) T_A({\bf{s}}))}\right] \nonumber \\
\times \left[ 1 -
{\exp(- \sigma_{abs} (1 - {1/B}) T_B({\bf{s-b}}))}\right].
\end{eqnarray}
Here b is the impact parameter of collision and s describes the 
transverse coordinates of the interacting nucleons which produce 
c$\bar{c}$.
This equation can be written in a more simpler form, 
\begin{equation}
S_{AB} = \exp(- \sigma_{abs} \rho (L_A + L_B)),
\end{equation}
where $L_A$+$L_B$ is the effective length a $c\bar{c}$ travels inside
two nuclei. 
Except for the case of Pb-Pb
collision, where an anomaluos $J/{\psi}$
suppression is observed by NA50 \cite{na50,na501},
this method
explains almost all datas upto $S-U$ collisions, using
a common absorption cross section $\simeq 6.2 mb$.

For a fixed impact parameter ${\bf{b}}$, this survival probability
is;
\begin{eqnarray}
S_{c\bar{c}-N}(b) = {1 \over {T_{A^{\prime}B^{\prime}}(b) 
{\sigma_{abs}}^2}} \int d^2 s \left[1 - {\exp(- \sigma_{abs}
(1 - {1/A}) T_A({\bf{s}}))}\right] \nonumber \\
\times \left[ 1 - {\exp(- \sigma_{abs} (1 - {1/B}) 
T_B({\bf{s-b}}))}\right],
\end{eqnarray}
where $T_{AB}(b) = \int d^2 s T_A({\bf{s}}) T_B({\bf{s-b}})$.

In our calculation we use a uniform density($R_A = r_0 A^{1/3}$, with 
$r_0 = 1.2 fm$), which is applicable
for heavy nuclei and have adopted a absorption cross
section of 6.2mb(used by the NA50 collaboration\cite{na501}).
For central Au-Au collisions($b = 0$), we obtain from the
above equation, $S_{c\bar{c}-N} \simeq 0.42$.

\section{results and discussions}

In fig-1 we present the thermal averaged
gluon-$J/{\psi}$ cross section
as a function of $J/{\psi}$ transverse momentum $P_T$, for different
values of temperature. It can be seen that this value is decreased
as the temperature of the plasma is increased.
Also for $J/{\psi}$ with high $P_T$ this thermal cross section
becomes less. This has a direct impact on $J/{\psi}$ suppression.
As $P_T$ becomes large the survival probability must increase.
However, due to large initial parton density formed in the
initial stage of the plasma, these smaller 
thermal cross sections do not have a large effects 
on the total survival probability.
This is shown in fig-2 where we have plotted 
the survival probability $S_{g-J/{\psi}}$ as a function of $P_T$
for a thermally equilibrating QGP at RHIC within PCM. 
As can be seen, the suppression is almost 100 percent
for $J/{\psi}$ with low transverse momentum.
Even for $J/{\psi}$ with high $P_T(\simeq 10 GeV)$
the suppression is about 95 percent. 
This huge suppression is due to the presence of a highly densed
deconfined partonic medium at RHIC.
As can be seen from equation(3) and fig-1 this suppression will be more
for a plasma with 
larger number density and lower temperature.
A denser and cooler plasma is obtained once the soft partons are
added to the hard and semi hard partons \cite{nayak1}.
The measurement of such huge 
suppression will reveal the existence of such a deconfined partonic 
medium, possibly a thermalised quark-gluon plasma, at RHIC.

However, what one actually measures experimentally is the total survival
probability. This implies that other
sources of $J/{\psi}$ suppression, without the existence
of a QGP phase, have to be 
singled out from the total survival probability in order to
see the suppressions only from this deconfined partonic medium.
In secton III,
we have estimated a large suppression (about 60 percent) due to 
the nuclear abosorption before a $J/{\psi}$ is formed.
If the estimated 
suppression due to all sources without QGP phase is much
less than the experimental findings, one may then hope 
that the rest of the suppressions
is due to the existence of a highly densed deconfined partons.
The experimental measurement of $J/{\psi}$ at RHIC will 
be able to reveal all the possibilities and may 
explain the existence of such a deconfined partonic
medium, possibly a thermalised quark-gluon plasma.

On the other hand $J/{\psi}$ enhancement can occur 
due to an increase in the (thermal)charms produced at a very
high temperature, especially in the initial stage of the evolution
of the system \cite{geiger1,shor}. We recall that
screening of $J/{\psi}$ was originally \cite{matsui} 
proposed without taking production of thermal charms into account,
which was argued to be negligible at $T \simeq$ 200 - 300 MeV.
However, the system may approach a temperature which is as large as  
900 MeV initially, and can therefore 
produce a large number of thermal charms at RHIC. So to make resonable
estimates at RHIC and LHC,
one has to take into account both thermal and hard $J/{\psi}$.
For simplicity, we have not considered
here those $J/{\psi}$ which are created
from thermal charms.

\section{conclusion}

We have found a reasonable suppression(around 60 percent)
before a $J/{\psi}$ is fully formed, due to the 
nuclear absorptoin alone and an additional
suppression of about 95-100 percent due to a deconfined phase. 
These two mechanisms give a huge suppression, almost 100 percent,
to those $J/{\psi}$ which are produced 
from primary charm quarks.
In principle, the total survival probability is written as 
\begin{equation}
S_T = S_{c\bar{c}-N} S_{g-{J/{\psi}}} S_{other}
\end{equation}
where first two suppressions ($S_{c\bar{c}-N}$, $S_{g-J/{\psi}}$)
are due to pre-resonance nuclear absorption 
and g-${J/{\psi}}$
dissociation, and 
the last one($S_{other}$) is due to any other possible sources
of suppression \cite{gavin1,wong,hwa} such as 
final state comover scatterings. 
As the hadron-$J/{\psi}$ 
cross section is smaller than g-$J/{\psi}$ cross section \cite{hwa2},
we expect a lesser suppression of $J/{\psi}$ in the
hadronisation stage. 
The prominent suppression seems to be due to the nuclear
absorption and $J/{\psi}$ dissociation in the deconfined partonic
system. Hence the measurement of a huge $J/{\psi}$ suppression at
RHIC may
suggest the existence of a partonic plasma.
However, many refinements, such as space time
evolution of quarkonium production in nucleus-nucleus collision
using NRQCD has to be worked out, before
unambigous conclusions can be drawn for the pre-resonance suppression
of $J/{\psi}$.
In any case the contribution from thermal
$J/{\psi}$, which may not be neglible at RHIC and LHC, has to be
taken into account. It is only after incorporating the above features
the certainty of screening can be justified.

\vspace{2.1cm}

{\bf Acknowlegements:}

		 I am thankul to V. Ravishankar and P K. 
Jain for various discussions  I had with them.

\vspace{5.0cm}

\subsection*{\bf {Figure captions}}
\noindent
{\bf FIG.~1.} The thermal averaged gluon-$J/{\psi}$ cross section
$<v_{rel} \sigma>$ as a function of transverse momentum $P_T$ at
different temperatures. Solid line refers to T= 0.2 GeV, upper broken
line corresponds to T= 0.4 GeV and lower broken line refers to T=0.8 GeV.

\noindent
{\bf FIG.~2.} The survival probability of $J/{\psi}$ in a
thermally equilibrating plasma at RHIC.


\begin{thebibliography}{99}
\bibitem{nayak1} A minijet source term is added to
       color flux-tube model \cite{nayak}.
       The transport equation is solved in SU(3) by taking
       the quark-antiquark
       and gluon production via non-perturbative Schwinger mechanism 
       and by the perturbative minijets 
       to see the complete evoluion of quark-gluon plasma.
\end{thebibliography}
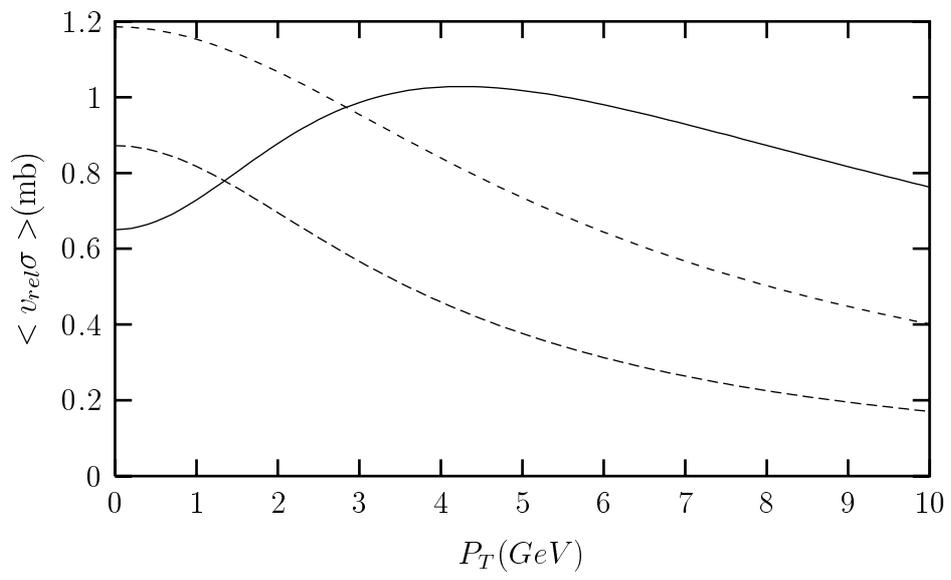
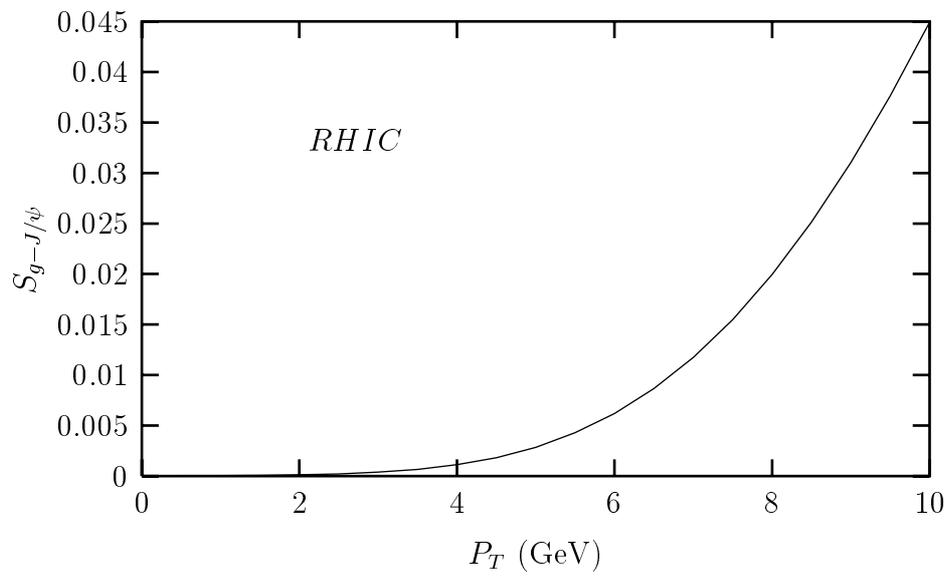
\end{document}